\documentstyle[aps,twocolumn,epsfig]{revtex} 
\begin{document}
{\bf  Comment  on  ``Stripe Glasses:  Self-Generated   Randomness in a
Uniformly Frustrated System''}\\

In  a recent letter, Schmalian  and  Wolynes\cite{SW00} have studied a
uniformly frustrated system whose Hamiltonian is given by
\begin{eqnarray}
\label{eq:ham1}  
\mathcal{H}&=&\frac{1}{2}\int d^3x\{r_0\phi({\bf{x}})^2+[\nabla\phi({\bf{x}})]^2
+\frac{u}{2}\phi({\bf{x}})^4\}\nonumber\\
&+&\frac{Q}{2}\int d^3x\int d^3x'\frac{\phi({\bf{x}})
\phi({\bf{x'}})}{|{\bf{x}}-{\bf{x'}}|}.
\end{eqnarray}
Using  the replica    formalism  and  the   self-consistent  screening
approximation,  they  show that  the  competition  of  interactions on
different length scales leads, below a crossover temperature $T_A$, to
the emergence of an  exponentially  large number of metastable  states
and, at a lower temperature  $T_k$, to a phase  transition to a glassy
state.   Moreover, from entropic droplet  arguments  they predict that
slow activated dynamics should occur at temperatures between $T_A$ and
$T_K$, with  the relaxation  time  $\tau$ obeying  a  Vogel-Fulcher law,
$\tau\propto\exp(\frac{DT_k}{T-T_k})$, and the  fragility parameter $D$ being
proportional to $(\frac{\partial  S_c}{\partial T}|_{T_k})^{-1}$,  where $S_c(T)$,
the  configurational entropy,  is  the  logarithm   of  the number  of
metastable states.   Since  they   find  that $\frac{\partial   S_c}{\partial  T}$
decreases   when the frustration  parameter $Q$  decreases, the system
should  become  less  fragile (i.e.,  with  a larger  $D$)  when  $Q$
decreases.   Such   a  conclusion  is  strikingly  at   odds  with the
prediction made for  similar systems by the frustration-limited domain
theory of the glass transition\cite{KKZNT95}.

\begin{figure}\vspace*{-0.5cm}
\begin{center}

\resizebox{8cm}{!}{\includegraphics{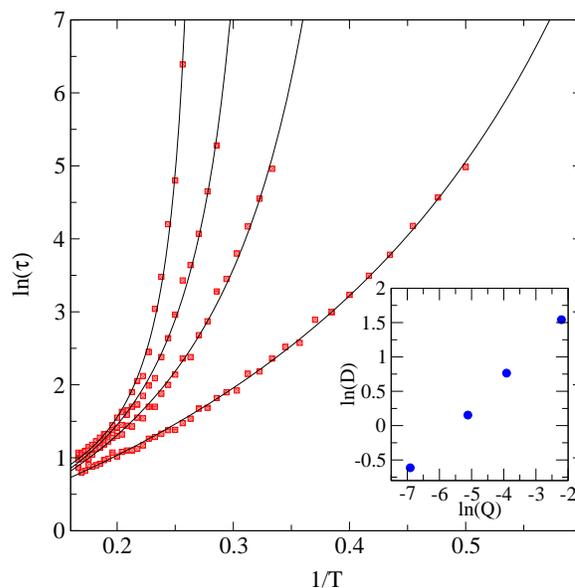}}
\caption{$\ln(\tau)$ versus $1/T$ for  $Q=0.001,0.006,0.02,0.11$    (from left  to
right). Solid  lines:  fits to  the  Vogel-Fulcher  law. The higher  T
values (down to $\ln(\tau) \sim 0$) are used in the fit, but are not shown
here. Inset: $\ln(D)$ versus $\ln(Q)$.}\label{fig:1}
\end{center}\end{figure}

We    comment  here   on  the   connection   made  by   Schmalian  and
Wolynes\cite{SW00}  between   the  configurational  entropy    and the
relaxational behavior  of the  frustrated  system and on  the relation
between  the fragility of a  glass-forming system and the frustration.
We have carried out computer  simulations of the ``hard-spin'' lattice
version  of   the  field-theoretical  action  in  Eq.~(\ref{eq:ham1}),
namely,
\begin{equation}
\label{eq:ham2}
H=-\sum_{<i,j>}S_iS_j+\frac{Q}{2}\sum_{i\neq j}\frac{S_iS_j}{r_{ij}},
\end{equation}
where the spins, $S_i=\pm1$,  are placed on a cubic  lattice\cite{VT98}.
By using the Metropolis  algorithm with the  constraint of  zero total
magnetization,    we  have    computed the    (equilibrium)  spin-spin
correlation   function,  $C(t)=\frac{1}{N}\sum_i<S_i(0)S_i(t)>$,  as   a
function of temperature for a range  of frustration parameter $Q$ that
covers  the values studied  in  Ref.\cite{SW00}.  The relaxation  time
$\tau$    has been   obtained    from  the  simulation  in   a  standard
way\cite{NC98} as  the  time  at which $C(t)=0.1$.    The  results are
reported in Fig.\ref{fig:1}.

Although   other formulas can be  used  as well\cite{KKZNT95}, we have
fitted  our  simulation  data to   the Vogel-Fulcher law  discussed in
Ref.  \cite{SW00}, and, as seen from   Fig.  \ref{fig:1}, the fits are
very good  for  all  values of  $Q$.   In the   inset, we display  the
fragility parameter $D$  versus  $Q$ on  a  $ln-ln$ plot: $D$  roughly
increases  as  $\sqrt{Q}$.  Clearly, the   uniformly frustrated system
becomes less fragile when the  frustration $Q$ increases, as can  also
be  seen from    the curvature of  the   various  $ln(\tau)$ curves   in
Fig.\ref{fig:1}.  This result, that fragility decreases as frustration
increases, disagrees with the  analysis presented  in Ref.\cite{SW00},
but supports the prediction  of the frustration-limited  domain theory
\cite{KKZNT95}.

The above  discussion seems to  suggest that, contrary to the commonly
held view, the  relaxation time of  a system that possesses a complex,
rugged free-energy  landscape (which, as shown  in \cite{SW00}, is the
case   of the uniformy frustrated system) is    not solely, nor primarily,
determined by the number of available  metastable states, i.e., by the
configurational  entropy.   Other      ingredients (preferred   paths,
free-energy barriers, connectivity of the  minima) may be necessary as
well.

\vspace*{-0.5cm}

M. Grousson, G. Tarjus and P. Viot
\\
Laboratoire de  Physique Th{\'e}orique des  Liquides,  Universit{\'e}
Paris VI,   4,   place Jussieu  75252 Paris  Cedex  05
France



\vspace*{-0.5cm} 
\begin{thebibliography}{1}
\bibitem{SW00}
J. Schmalian and P. Wolynes, Phys. Rev. Lett. {\bf 85},  836  (2000).
\bibitem{KKZNT95}
D. Kivelson {\it et al.}, Physica A {\bf 219}, 27 (1995);  G. Tarjus
{\it et al.} J. Phys.: Cond. Matter {\bf 12}, 6497 (2000).
\bibitem{VT98}
P. Viot and G. Tarjus, Europhys. Lett {\bf 44},  423  (1998).
\bibitem{NC98}
M. Nicodemi and A. Coniglio, Phys. Rev. E {\bf 57}, R39 (1998).



\end{thebibliography}
\end{document}